\documentclass[pra,twocolumn,showpacs,nofootinbib,british,amsmath,amssymb,draft]{revtex4}

\usepackage[british]{babel} %\usepackage{latexsym}\usepackage{textcomp}
\usepackage{bm}
\newcommand{\kron}{\mathrm{\delta}}%Kronecker

\renewcommand{\implies}{\Rightarrow}%implies
\newcommand{\st}{\mid}%such that
\newcommand{\with}{\mathpunct{:}}%with (list of indices)
\newcommand{\defin}{\stackrel{\text{def}}{=}}%definition
\renewcommand{\le}{\leqslant}%less or equal
\renewcommand{\ge}{\geqslant}%greater or equal
\newcommand{\reals}{\mathbb{R}}
\DeclareMathOperator{\tr}{tr}%trace
\DeclareMathOperator{\rank}{rank}%rank
\DeclareMathOperator{\card}{card}%cardinality
\newcommand{\inn}{\cdot}%inner product
\newcommand{\tran}{\mathsf{T}}%transpose
\newcommand{\set}[1]{\{#1\}}

\newcommand{\settT}[1]{\Biggl\{#1\Biggr\}}
\newcommand{\ie}{\textit{i.e.}}
\newcommand{\eg}{\textit{e.g.}}
\newcommand{\viz}{\textit{viz.}}
\newcommand{\cf}{\textit{cf.}}

\newcommand{\zM}{M}%no. states
\newcommand{\zss}{s}%states
\newcommand{\zS}{s}%state comp
\newcommand{\zs}{\bm{s}}%state vect
\newcommand{\zSS}{S}%state set
\newcommand{\zSC}{S_{\text{c}}}%state set's conv hull
\newcommand{\zm}{m}%measurements
\newcommand{\zrr}{r}%outcomes
\newcommand{\zR}{r}%outcome comp
\newcommand{\zr}{\bm{r}}%outcome vect
\newcommand{\zRR}{R}%outcome set
\newcommand{\zRC}{R_{\text{c}}}%outcome set's conv hull
\newcommand{\zL}{L}%no.outcomes
\newcommand{\zP}{\bm{p}}%table
\newcommand{\zp}{p}%probabilities
\newcommand{\zK}{K}%rank
\newcommand{\zT}{\bm{t}}% outcome table
\newcommand{\zt}{t}%outcome table entry
\newcommand{\zU}{\bm{u}}%state table
\newcommand{\zu}{u}%state table entry
\newcommand{\zA}{\bm{a}}% submatrix
\newcommand{\zB}{\bm{b}}% submatrix
\newcommand{\zC}{\bm{c}}% submatrix
\newcommand{\zD}{\bm{d}}% submatrix
\newcommand{\zV}{\bm{v}}% submatrix
\newcommand{\zW}{\bm{w}}% submatrix
\newcommand{\zX}{\bm{x}}% submatrix
\newcommand{\zY}{\bm{y}}% submatrix
\newcommand{\zI}{I}% set of indices
\newcommand{\zd}{D}% num of outcomes
\newcommand{\zn}{\bm{n}}% sum of outcome vects
\newcommand{\ze}{\bm{q}}% matrix (11111...)
\newcommand{\zla}{\lambda}% convex coeff 1
\newcommand{\zmu}{\mu}% convex coeff 2
\newcommand{\zE}{\bm{E}}% id matrix
\newcommand{\zF}{\bm{F}}% map
\newcommand{\zg}{\bm{g}}% map
\newcommand{\zc}{\bm{C}}% lin map
\newcommand{\zN}{N}% max no. disting. states
\newcommand{\zx}{Z}% no. extreme points
\newcommand{\zo}{\bm{o}}% origin
\newcommand{\zrho}{\Hat{\bm{\rho}}}% dens matr
\newcommand{\zpov}{\Hat{\bm{A}}}% POV el
\newcommand{\zbas}{\Hat{\bm{B}}}% basis
\newcommand{\zSQ}{\zSS_{\text{QM}}}% QM states
\newcommand{\zRQ}{\zRR_{\text{QM}}}% QM states
\allowdisplaybreaks[1]

\begin{document}
\title{Why can states and measurement outcomes be represented as vectors?}

\author{Piero~G.~L.~Mana}

\email{mana@imit.kth.se}

\affiliation{Institutionen f\"or Mikroelektronik och
  Informationsteknik (IMIT),\\
%Department of Microelectronics and Information Technology,\\
%Kungliga Tekniska H\"ogskolan\\
Royal Institute of Technology (KTH),\\
Electrum 229, SE-164\,40 Kista, Sweden}

\date{20 May 2003}

\begin{abstract} 
It is shown how, given a ``probability data table'' for a
quantum or classical system, the representation of states and
measurement outcomes as vectors in a real vector space follows
in a natural way. Some properties of the resulting sets of these
vectors are discussed, as well as some connexions with the
quantum-mechanical formalism.
\end{abstract}


\maketitle

\section{Introduction}
\label{sec:intro}

It has long been known that quantum mechanics is expressible through
formalisms which differ, more or less, from the classical
complex-Hilbert-space-based one; examples where given, among others,
by Wigner~\cite{wigner1932}, Stapp~\cite{stapp1971},
Wootters~\cite{wootters1986,wootters1987},
Weigert~\cite{weigert2000}, Hardy~\cite{hardy2001},
Havel~\cite{havel2003}. Some of these formalisms may be more useful
than others in practical applications, but certainly all are very
useful to understand better the physics ``behind'' quantum mechanics.
This is particularly true when their mathematics is simple or, for
example, geometrically appealing.

This is certainly the case for Hardy's formulation~\cite{hardy2001}
of quantum mechanics in the simple language of real vector spaces.

Hardy's main idea, which was already expressed by
Peres~\cite{peres1995}, is that a system's \emph{state}, or
\emph{preparation}, is characterised by a list of ``all probabilities
for all measurements that could possibly be performed'' on the
system. However, such a list of probabilities is likely to be
(infinite and) over-complete, ``since most physical theories have some
structure which relates different measured quantities''; discarding
the ``redundant'' probabilities from the list, all that is left is
simply a vector of real numbers: this is the state. An analogous
conclusion can be drawn for the representation of \emph{measurement
outcomes}.

Such a framework is very general, and not restricted to the
description of states and outcomes of quantum-mechanical systems
only. Hardy characterises the latter by means of some simple,
``reasonable'' axioms.

A similar idea is proposed in the present paper, but from a slightly
different perspective. It is shown that, given an experimental `data
table', containing statistical data about a generic classical or
quantum system, the representation of preparations and measurement
outcomes as vectors in a real vector space follows naturally as an
alternative way of organising, or storing, the table's data. Some
properties of the sets of these vectors are studied. The framework is
general, not restricted to quantum mechanics; connexions with the
latter are discussed in the end.

\section{Decomposition of a data table}\label{sec:decom}
Consider a system, which can be classical, quantum, or of unknown
nature. It is easy to imagine the following: A physicist can prepare
this system in a given number $\zM$ of different \emph{preparations},
or \emph{states} $\set{\zss_1, \dotsc, \zss_{\zM}}$, and perform on
it a given number of measurements $\set{\zm_k}$, each with a
different number of \emph{outcomes} $\set{\zrr_i \with i\in
\zI_{\zm_k}}$ (mutually exclusive and
exhaustive%
\footnote{This can always be achieved by grouping in suitable ways
the outputs of the measurement, and adding if necessary the outcome
``\emph{other}''.}), where the sets of indices $\zI_{\zm_k}$ depends
on the measurement in question. The total number of outcomes of all
measurements is $\zL$. Through experiments (and, possibly but not
necessarily, some theoretical reasoning), the physicist can write
down a table $\zP$ with the probabilities for every outcome, for
every measurement and state prepared; it may look like the following:
%Table~\ref{tab:table}.

\begin{table}[!h]%\caption{\label{tab:table}Probability table}
\begin{tabular}{cc|cccccc}
% &\multicolumn{6}{c}{states}
&  & $\zss_1$& $\zss_2$& $\zss_3$& 
$\zss_4$& $\dotso$& $\zss_{\zM}$ \\
\hline
$\zm_1$&$\zrr_1$ & $\zp_{11}$& $\zp_{12}$& $\zp_{13}$& 
$\zp_{14}$&$\dotso$& $\zp_{1\zM}$ \\
& $\zrr_2$ & $\zp_{21}$& $\zp_{22}$& $\zp_{23}$& 
$\zp_{24}$&$\dotso$& $\zp_{2\zM}$ \\
\cline{1-2}
$\zm_2$&$\zrr_3$ & $\zp_{31}$& $\zp_{32}$& $\zp_{33}$& 
$\zp_{34}$&$\dotso$& $\zp_{3\zM}$ \\
& $\zrr_4$ & $\zp_{41}$& $\zp_{42}$& $\zp_{43}$& 
$\zp_{44}$&$\dotso$& $\zp_{4\zM}$ \\
& $\zrr_5$ & $\zp_{51}$& $\zp_{52}$& $\zp_{53}$& 
$\zp_{54}$&$\dotso$& $\zp_{5\zM}$ \\
\cline{1-2}
$\zm_3$& $\zrr_6$ & $\zp_{61}$& $\zp_{62}$& $\zp_{63}$& 
$\zp_{64}$&$\dotso$& $\zp_{6\zM}$ \\
&$\dotso$& \multicolumn{6}{c}{$\dotso$}\\
&  $\zrr_{\zL}$ & $\zp_{\zL 1}$& $\zp_{\zL 2}$& $\zp_{\zL 3}$& 
$\zp_{\zL 4}$&$\dotso$& $\zp_{\zL\zM}$ \\
\end{tabular}
\end{table}

The table, which can be called a `\emph{(experimental) probability
data table}' or `data table' for short, has a column for every state
and a row for every outcome; the table entry $(i,j)$ (\eg, $(4,2)$)
is the probability $\zp_{ij}$ ($\zp_{4\,2}$) of obtaining the outcome
$\zrr_i$ ($\zrr_4$, among the possible outcomes
$\set{\zrr_3,\zrr_4,\zrr_5}$ of the measurement $\zm_2$) when the
system is prepared in the state $\zss_j$ ($\zss_2$). States and
outcomes can be listed and rearranged in any desired way in the
table. Such a table would very likely have a large number of rows and
columns, \ie, the numbers $\zL$ and $\zM$ are likely to be very
large.\footnote{Of course, one may wonder how often a physicist has
to concretely deal with similar tables, or whether a similar table has
ever been written down actually; yet, it cannot be denied that this
imaginary table conveys an idea of a part of that complex activity
called ``doing physics''. Its main purpose here is to give a
completely operational background to the concepts presented. It must
also be remarked that it is not strictly necessary to speak about
systems, states or preparations, and measurement outcomes: a similar
table could be compiled by considering how different aspects of a
given phenomenon are (cor)related to each other in their various
manifestations. The terms `system', `state', `measurement', and
`outcome' will nevertheless be used here for definiteness.}

Suppose that the physicist wants to find a more compact, or simply
different, way to write down and store the data in $\zP$. The table
$\zP$ is really just an $\zL \times \zM$ rectangular matrix, and as
such it has a rank $\zK$, \viz, the minimum number of linearly
independent rows or columns:
\begin{equation}\label{eq:rank}
\zK\defin\rank\zP \le \min\set{\zL,\zM}.
\end{equation}
It follows from linear algebra that $\zP$ can be written as the
product of an $\zL \times \zK$ matrix $\zT$ and a $\zK \times \zM$
matrix $\zU$%
\footnote{This is equivalent to the fact that a linear map
$\zp\colon\reals^{\zM} \to \reals^{\zL}$ of rank $\zK\defin\dim
\zp\bigl(\reals^{\zM}\bigr)$ can be obtained as the composition
$\zp=\zt\circ\zu$ of a surjective map $\zu\colon \reals^{\zM} \to
\zp\bigl(\reals^{\zM}\bigr)$ and an injective map $\zt\colon
\zp\bigl(\reals^{\zM}\bigr) \to \reals^{\zL}$.}:
\begin{equation}
\label{eq:decomp1}
\zP =\zT\,\zU,
\end{equation}
or
\begin{multline}
\left(\begin{smallmatrix}
\zp_{11} & \dots &\zp_{1j} &\dots&\zp_{1\zM} \\
\dots&\dots&\dots&\dots&\dots\\
\zp_{i1} & \dots &\zp_{ij} &\dots&\zp_{i\zM} \\
\dots&\dots&\dots&\dots&\dots\\
\zp_{\zL 1} & \dots &\zp_{\zL j} &\dots&\zp_{\zL\zM}
\end{smallmatrix}\right) = \\
\begin{aligned}
&= \left(\begin{smallmatrix}
\zt_{11} & \dots &\zt_{1\zK} \\
\dots&\dots&\dots\\
\zt_{i1} & \dots&\zt_{i\zK} \\
\dots&\dots&\dots\\
\zt_{\zL 1} & \dots &\zt_{\zL\zK}
\end{smallmatrix}\right)
\left(\begin{smallmatrix}
\zu_{11} & \dots &\zu_{1j} &\dots&\zu_{1\zM} \\
\dots&\dots&\dots&\dots&\dots\\
\zu_{\zK 1} & \dots &\zu_{\zK j} &\dots&\zu_{\zK\zM}
\end{smallmatrix}\right)
\\
&= \left(\begin{smallmatrix}
\zr_1^{\tran} \\
\dots\\
\zr_i^{\tran} \\
\dots\\
\zr_{\zL}^{\tran} 
\end{smallmatrix}\right) 
\left(\begin{smallmatrix}
\zs_1 & \dots &\zs_j &\dots&\zs_{\zM} \\
\end{smallmatrix}\right) 
\end{aligned}
\end{multline}
In the last equation, the matrix $\zT$ has been written as a block of
row vectors $\zr_i^{\tran}$, and the matrix $\zU$ as a block of
column vectors $\zs_i$. In this decomposition, the element $\zp_{ij}$
of $\zP$ is then given by the matrix product of the row vector
$\zr_i^{\tran}$ with the column vector $\zs_j$:
\begin{equation}
\label{eq:baserule}
\zp_{ij}=\zr_i^{\tran} \zs_j=\zr_i \inn \zs_j,\qquad
\zr_i,\zs_j\in\reals^{\zK},
\end{equation}
where, in the last expression, the row $\zr_i$ and $\zs_j$ are
considered as vectors in $\reals^{\zK}$, so that the matrix product
is equivalent to the scalar product.

The result is that, given a probability table for a system, the
states and the measurement outcomes for that system can be
represented as \emph{vectors}, $\set{\zs_j}$ and $\set{\zr_i}$, in
$\reals^{\zK}$, for some $\zK$, and the relative probabilities are
given by their \emph{scalar product}. These vectors can be called
\emph{state vectors} and \emph{outcome vectors}. As already said,
this kind of representation has already been presented by
Hardy~\cite{hardy2001} through a different line of reasoning. In
particular, Hardy supposes that it is possible to represent a state
with a $\zK$-dimensional vector\footnote{Hardy calls $\zK$ the
\emph{number of degrees of freedom} of the system.}, with
$\zK\le\zL$, because ``most physical theories have some structure
which relates different measured quantities''; but the reasoning
above shows that this possibility exists even before building up some
theory to describe the data.

The matrices $\zT$ and $\zU$ are not uniquely determined from the
decomposition~\eqref{eq:decomp1}, so that there is some freedom in
choosing their form. The fact that $\rank\zP=\zK$, implies that there
exists a square $\zK\times\zK$ submatrix $\zA$, obtained from $\zP$
by suppressing $(\zL-\zK)$ rows and $(\zM-\zK)$ columns, such that
$\det\zA\neq 0$. It is always possible to rearrange the rows and the
columns of the table $\zP$ so that such submatrix is the one formed
by the first $\zK$ rows and $\zK$ columns. After this rearrangement
(which, of course, does not imply any physical operation on the
system), $\zP$ can be written in the following block form:
\begin{equation}
\label{eq:blockformA}
\zP=\begin{pmatrix}\zA&\zB\\ \zC & \zD \end{pmatrix}\quad
\text{with $\det\zA\neq 0$},
\end{equation}
where $\zB$, $\zC$, and $\zD$ are of order $\zK\times(\zM-\zK)$,
$(\zL-\zK)\times\zK$, and $(\zL-\zK)\times(\zM-\zK)$ respectively.

By writing also the matrices $\zT$ and $\zU$ in block form
\begin{equation}
\label{eq:blockformTU}
\zT=\begin{pmatrix}\zV\\ \zW \end{pmatrix},\quad
\zU=\begin{pmatrix}\zX & \zY \end{pmatrix},
\end{equation}
where the orders of $\zV$, $\zW$, $\zX$, and $\zY$ are
$\zK\times\zK$, $(\zL-\zK)\times\zK$, $\zK\times\zK$, and
$\zK\times(\zM-\zK)$ respectively, we can rewrite
Eq.~\eqref{eq:decomp1} as
\begin{equation}\label{eq:blockdecomp}
\begin{pmatrix}\zA&\zB\\ \zC & \zD \end{pmatrix} =
\begin{pmatrix}\zV\\ \zW \end{pmatrix} 
\begin{pmatrix}\zX & \zY \end{pmatrix},\quad\text{or}\quad
\left\{\begin{aligned}
\zA&=\zV \zX \\
\zB&=\zV \zY \\
\zC&=\zW \zX \\
\zD&=\zW \zY
\end{aligned}\right.
\end{equation}
with the solution\footnote{The submatrix $\zD$ of $\zP$ is completely
determined by the other submatrices $\zA$, $\zB$, and $\zC$ because
$\rank\zP=\zK$.}
\begin{equation}\label{eq:solutiondecomp}
\left\{\begin{aligned}
\zV&=\zA \zX^{-1}\\ \zW&=\zC \zX^{-1} \\ 
\zY&=\zX \zA^{-1} \zB \\ \zD&=\zC \zA^{-1} \zB
\end{aligned}\right.\quad\text{or}\quad
\left\{\begin{aligned}
\zT&=\begin{pmatrix}\zA \zX^{-1}\\ \zC \zX^{-1} \end{pmatrix}\\
\zU&=\begin{pmatrix}\zX & \zX \zA^{-1} \zB \end{pmatrix}
\end{aligned}\right.\quad \det\zX\neq 0,
\end{equation}
where the square matrix
\begin{equation}\label{eq:matrixX}
\zX = (\zs_1\;\dots\;\zs_{\zK})
\end{equation}
is undetermined except for the condition of being non-singular; this
corresponds to the freedom of choosing $\zK$ \emph{basis vectors} in
$\reals^{\zK}$ as the representatives of the first $\zK$
states\footnote{There is the alternative option of choosing the
representatives of the first $\zK$ outcomes; this corresponds to
solving Eq.~\protect\eqref{eq:blockdecomp} in terms of $\zV$.}
$\set{\zs_1,\dotsc,\zs_{\zK}}$, which can then be called \emph{basis
states}.

\section{The sets of states and outcomes}
\label{sec:sets}
The vectors by which states and outcomes are represented belong to
two subsets, $\zSS\defin\set{\zs_1,\dotsc,\zs_{\zM}}$ and
$\zRR\defin\set{\zr_1,\dotsc,\zr_{\zL}}$ respectively, of
$\reals^{\zK}$. It is interesting to study some properties of these
sets. Some of the following results have been obtained by Peres and
Terno~\cite{peresetal1998} in the framework of quantum mechanics. The
convex-related properties of the sets are also well known~\cite[and
references therein]{buschetal1984,busch1986,barnum2003}, but are often
derived and expressed through more elegant, and abstract,
mathematics.

In the following, the terms `vector' and `point' are used
interchangeably.

\subsection{``Completion'' of the sets}
\label{sec:completionsets}
One can consider the convex hull of the set of states $\zSS$:
\begin{equation}\label{eq:convhullstates}
\zSC\defin \settT{\sum_{i=1}^{\zM}\zla_i \zs_i \st
\zs_i\in\zSS,\;\zla_i\ge0,\;\sum_{i=1}^{\zM}\zla_i=1}.
\end{equation}
An element of $\zSC$ like, \eg, $\zla \zs_{j'}+(1-\zla)\zs_{j''}$,
with $\zs_{j'},\zs_{j''}\in\zSS$ (and $0\le\zla\le 1$), can be
considered as a possible state, corresponding to a preparation in
which $\zs_{j'}$ or $\zs_{j''}$ are chosen with probabilities $\zla$
or $(1-\zla)$ respectively. This state should then be added to the
table $\zP$, with a respective column of probabilities. However, this
column would be, for obvious reasons, just a linear combination of
the columns under $\zs_{j'}$ and $\zs_{j''}$, with coefficients
$\zla$ and $(1-\zla)$; as a consequence, the rank of the table would
be still $\zK$. For this reason, the set $\zSS$ and its convex hull
$\zSC$ can be used interchangeably in the considerations to
follow\footnote{This corresponds to ``completing'' the table $\zP$
with this (infinite) number of ``additional'' states; so $\zM$ tends
to infinity, but the rank $\zK$ remains constant.}. Reasoning in
terms of convexity, one sees that $\zSC$ is just a convex (not
necessarily regular) polytope in a $(\zK-1)$-dimensional (Euclidean)
space (for example, a triangle, rhomboid, or general polygon in two
dimensions; or a tetrahedron, cube, prism, or general polyhedron in
three, or a 600-cell in four, and so on~\cite{coxeter1948}). Some of
the states $\set{\zs_1, \dotsc, \zs_{\zM}}$ are \emph{extreme} points
of this convex set (vertices of the polytope). Since every state
vector can be written as a convex, hence linear, combination of these
\emph{extreme states}, they must be at least as numerous as the basis
states. The number $\zx$ of extreme states (in quantum mechanics,
they are called \emph{pure} states) must then satisfy
% $\zx$ is always finite in practice (\ie, only this case may appear in
% an actual probability data table), though the limits of a continuous
% set of extreme states (or of infinite $\zK$, or both) may of course
% be considered.
\begin{equation}
\zx\ge\zK.\label{eq:ZK}
\end{equation}

For the set of outcomes, the situation is slightly different, and
convex combination is not the only way in which outcomes can be
combined. Consider, as a concrete example, the state $\zs_j$ and the
measurements $\zm'$, $\zm''$, and $\zm'''$ with outcomes
$\set{\zr_1,\zr_2,\zr_3}$, $\set{\zr_4,\zr_5}$, and
$\set{\zr_6,\zr_7}$ respectively, and the corresponding
probabilities; from them, the following additional measurements can
be derived:
\begin{itemize}
\item The measurement which consists in performing $\zm'$ but
considering only the set of two results $\set{(\zr_1\text{ or
}\zr_3), \zr_2}$, with probabilities $\set{(\zp_{1j} + \zp_{3j}),
\zp_{2j}}$ (a sort of coarse-graining).
\item The measurement which consistes in performing $\zm'$ with
probability $\zla'$, or $\zm''$ with probability $\zla''$, or
$\zm'''$ with probability $\zla'''$ (obviously,
$\zla'+\zla''+\zla'''=1$). The experimenter (who may not know which
measurement will actually be performed) expects thus one of the
outcomes $\set{\zr_1,\zr_2,\zr_3,\zr_4,\zr_5,\zr_6,\zr_7}$ with
probabilities $\set{\zla'\zp_{1j},\zla'\zp_{2j}, \zla'\zp_{3j},
\zla''\zp_{4j}, \zla''\zp_{5j}, \zla'''\zp_{6j}, \zla'''\zp_{7j}}$
(of course, obtaining, \eg, the result $\zr_2$ would imply that
$\zm'$ was actually performed).
\item Combinations of the two cases above.
\end{itemize}
From the examples just given, it is easy to see that, given two
outcomes $\zr_{i'}$ and $\zr_{i''}$, not necessarily of the same
measurement, one can consider also the outcomes
\begin{gather}
\zla' \zr_{i'} + \zla'' \zr_{i''}\quad\text{($0\le \zla'+\zla''\le 1$
and $\zla', \zla''\ge 0$)},\label{eq:comboutc1}\\
\intertext{and} 
\zr_{i'}+\zr_{i''}\quad\text{(only if $\zm' = \zm''$)}.\label{eq:comboutc2}
\end{gather}
Note that Eq.~\eqref{eq:comboutc1} is \emph{not} (always) a convex
combination. The null vector (origin) $\zo$ belongs thus to the set
$\zRC$.

These measurements and outcomes could be added to the table $\zP$ as
well, with their relative rows of probabilities; the rank of the
table would nevertheless remain $\zK$ for the same reason given for
the additional states. Thus, also the set of outcomes $\zRR$ can be
ideally extended to a set $\zRC$ by means of
Eqs.~\eqref{eq:comboutc1} and~\eqref{eq:comboutc2}. Again, $\zRR$ and
$\zRC$ will be referred to interchangeably in the following.

\subsection{The set of states lies in a hyperplane}
\label{sec:stateshyperplane}
Consider a state $\zs_{j}\in\zSS$ and all the outcomes
$\set{\zr_{i'}\with i'\in \zI_{\zm'}}$ of a given measurement $\zm'$,
where the sets of indices $\zI_{\zm'}$ depends on $\zm'$. Since the
outcomes are exhaustive and mutually exclusive, their probabilities
must sum up to unity:
\begin{equation}\label{eq:sumunity}
\sum_{i'\in \zI_{\zm'}} \zp_{i'j} = 
\sum_{i'\in \zI_{\zm'}} (\zr_{i'}\inn \zs_{j})
=\Biggl(\sum_{i'\in \zI_{\zm'}} \zr_{i'}\Biggr)\inn \zs_{j}=1.
\end{equation}
The equation above, when considered for the first $\zK$ states
$\set{\zs_1,\dotsc,\zs_{\zK}}$, which form a basis for
$\reals^{\zK}$, uniquely determines the vector sum of the outcomes of
$\zm'$, denoted by $\zn'\defin \sum_{i'\in \zI_{\zm'}} \zr_{i'}$, by
its projections along the basis vectors. On the other hand, this
happens for the outcomes of any measurement $\zm$. Hence, the sum of
the outcome vectors $\set{\zr_{i}\with i\in \zI_{\zm}}$ of \emph{any}
measurement $\zm$ is a constant vector
\begin{gather}
\zn\equiv\sum_{i\in \zI_{\zm}} \zr_{i},\quad\text{for any
$\zm$},\label{eq:defnormal}\\
\intertext{such that}
\zn^{\tran}\, \zs \equiv \zn \inn \zs = 1.\label{eq:planeeq}
\end{gather}
The vector $\zn$ (which is an extreme point of $\zRC$) may be called
the \emph{trivial-measurement vector}, since it also represents the
outcome of the trivial measurement having only a single (and
therefore certain) outcome for any state. The actual components of
this vector are determined by the choice of the matrix $\zX$. In
fact, Eq.~\eqref{eq:planeeq} holds, in particular, for every basis
state $\set{\zs_1,\dotsc,\zs_{\zK}}$, and the set of these $\zK$
equations can be written, with the help of
formula~\eqref{eq:matrixX}, in the following matrix form:
\begin{equation}\label{eq:normaleqforbasis}
\zn^{\tran} \zX=\ze^{\tran},\quad
\text{with }\ze=
\overset{\text{$\zK$ elements}}{\bigl(\overbrace{\begin{matrix}1&
1& \dotso &1\end{matrix}}\bigr)}.
\end{equation}
Since $\zX$ is non-singular, one finds
\begin{equation}\label{eq:normaldef}
\zn^{\tran}=\ze^{\tran} \zX^{-1},
\end{equation}
so that $\zX$ determines $\zn$, as asserted.

Finally, the equation~\eqref{eq:planeeq}, where $\zn$ is now
considered constant, must be satisfied by every state $\zs$; this is
the equation, in vector form, of a (affine) hyperplane normal to
$\zn$. Hence, the set of states $\zSS$ lies in a
$(\zK-1)$-dimensional (affine) hyperplane in $\reals^{\zK}$.

\subsection{The set of states determines the maximal possible
extension of the set of outcomes}
\label{sec:boundaryconstr}

The sets $\zSS$ and $\zRR$ can thus be viewed as compact, convex
regions in $\reals^{\zK}$. Their boundaries are
interrelated\footnote{They are \emph{dual}~\cite{buschetal1984}.}. In
particular, it is interesting to study how the boundary of $\zSS$
determines the region wherein $\zRR$ is constrained to lie (though
$\zRR$ may be a proper subset of this region).

Since $0 \le\zp\le 1$ for every probability $\zp=\zr \inn \zs$, one
has
\begin{equation}\label{eq:strips}
0 \le \zr \inn \zs \le 1.
\end{equation}
The formula above, with $\zs$ considered constant and $\zr$ variable,
defines a region in $\reals^{\zK}$ delimited by the parallel
hyperplanes $\zr \inn \zs =0$ and $\zr \inn \zs =1$. These
hyperplanes are both perpendicular to the vector $\zs$, and pass
respectively through the origin $\zo$ and through the point $\zn$,
since $\zn \inn \zs=1$ by Eq.~\eqref{eq:planeeq}. There is one such
region for every state $\zS\in\zSS$. The outcome vectors must be
confined to lie in the intersection of all these regions. However, if
an outcome vector $\zr$ lies inside the regions determined by two
states $\zs'$ and $\zs''$, it must also lie in that determined by any
state $\zla\zs'+(1-\zla)\zs''$ which is their convex combination;
\ie, if $0\le\zla\le1$, then
\begin{gather}\label{eq:striponlyextr}
\left.\begin{gathered}
0\le\zr\inn\zs'\le 1\\
0\le\zr\inn\zs''\le 1
\end{gathered}\right\}
\implies
0\le\zr\inn[\zla\zs'+(1-\zla)\zs'']\le 1.
\end{gather}
This is easily proven observing that, since $\zr\inn\zs'$,
$\zr\inn\zs''$, $\zla$, and $(1-\zla)$ are non-negative, then
$\zla\, \zr \inn \zs'+ (1-\zla)\, \zr \inn \zs''\ge0$; and since
$\zr\inn\zs',\zr\inn\zs''\le 1$, then $\zla\,
\zr\inn\zs'+(1-\zla)\,\zr\inn\zs''\le \zla +(1-\zla)=1$.

Hence, one needs to consider only the regions determined by the $\zx$
extreme states, which are the only ones not expressible as convex
combinations of other states. Now let
$\set{\zs_{i_1},\dotsc,\zs_{i_{\zx}}}$ be the extreme states.
Consider the $\zx$ hyperplanes of equations $\zr\inn\zs_{i_j}=0$,
$j=1,\dotsc,\zx$: they delimit a \emph{convex cone} with vertex in
the origin $\zo$. Another convex cone, with vertex in $\zn$, is
delimited by the other $\zx$ hyperplanes of equations
$\zr\inn\zs_{i_j}=1$, $j=1,\dotsc,\zx$. The intersection of these two
cones%
\footnote{It can be shown that the cones are symmetric with respect
to the point (center of symmetry) $\zn/2$: indeed, if
$\zr\inn\zs=0\text{ or }1$ for some $\zr$ and $\zs$, then
$[\zn/2+(\zn/2-\zr)]\inn\zs \equiv \zn\inn\zs-\zr\inn\zs =
1-\zr\inn\zs = 1\text{ or }0$, where $\zn/2+(\zn/2-\zr)$ is the point
symmetric to $\zr$ with respect to $\zn/2$.} finally determines the
maximal, $\zK$-dimensional, convex region which can be occupied by the set
of outcomes $\zRR$.
%  also that the only outcome vectors $\zr$ that lie on the
% boundary of $\zRR$ are those for which $\zr \inn \zs=0$ or $\zr \inn
% \zs =1$ for some state $\zs$.

\subsection{`One-shot' distinguishability}
\label{sec:disting}
Suppose that there exists a measurement $\zm$ that allows one to tell
with certainty which state, from a given set of $\zd$ states
$\set{\zs_{j_1}, \dotsc, \zs_{j_{\zd}}}$, is actually prepared. These
states can then be called `\emph{one-shot distinguishable}'. This
means that the table $\zP$ must contain a subtable of the form
(obtained, if necessary, by re-listing states and outcomes):

\begin{table}[h]
\begin{tabular}{cc|cccc}
% &\multicolumn{6}{c}{states}
& & $\zss_{i_1}$& $\zss_{i_2}$& $\dotso$& $\zss_{i_{\zd}}$ \\
\hline
$\zm$&$\zrr_{i_1}$ & $1$ & $0$& $\dotso$&$0$\\ 
&$\zrr_{i_2}$ & $0$ & $1$& $\dotso$&$0$\\ 
&$\dotso$& \multicolumn{4}{c}{$\dotso$}\\
 & $\zrr_{i_{\zd}}$ & $0$& $0$& $\dotso$&$1$
\end{tabular}
\end{table}

or, in other words, the matrix $\zP$ has a square $\zd\times\zd$
submatrix equal to the $\zd\times\zd$ identity matrix $\zE_{\zd}$.
This implies that 
\begin{equation}
\rank\zP\equiv\zK\ge\zd.\label{eq:KN}
\end{equation}

It can be proven that, if some of the states $\set{\zs_{j_1}, \dotsc,
\zs_{j_{\zd}}}$ are not extreme, they can be replaced by extreme
states having the same distinguishability
property~\cite[Sec.~6.12]{hardy2001}. Moreover, if the
$\set{\zs_{j_1}, \dotsc, \zs_{j_{\zd}}}$ are one-shot distinguishable
extreme states, then the convex combination of any $(\zd-1)$ of them
must lie on the boundary of $\zSS$; a simple proof of this fact for
the case of three one-shot distinguishable extreme states is given in
Appendix~\ref{sec:proofboundary}. This implies that, if $\zSS$
contains $\zd$ one-shot distinguishable states, there must exist a
$(\zd-1)$-dimensional hyperplane such that its intersection with
$\zSS$ is a $(\zd-1)$-dimensional simplex. 
% \footnote{In Hardy's
% axiomatics~\protect\cite{hardy2001}, quantum mechanics is
% characterised by (among other properties) the relation $\zK=\zN^2$,
% where $\zN$ is the maximum number of distinguishable states of the
% system; classical systems are instead characterised by $\zK=\zN$.}.

\emph{In particular}, this is also true for the set of states in
quantum mechanics. For example, the set of states of a three-level
system, which has at most three distinguishable states, shows
triangular two-dimensional sections~\cite[Fig.~2]{kimura2003}; for a
four-level system, a three-dimensional section of the set of states
yields a tetrahedron, and so some two-dimensional sections must have
triangular as well as trapezoidal
shapes~\cite[Fig.~1]{jakobczyketal2001}.

The \emph{maximum} number of one-shot distinguishable states for a
given system with table $\zP$ will be denoted by
$\zN$;\footnote{Hardy calls $\zN$ the \emph{dimension} of the
system.} from Eqs.~\eqref{eq:ZK} and~\eqref{eq:KN} it follows that
\begin{equation}
\zx \ge \zK \ge \zN\label{eq:relatZKN}.
\end{equation}

Note that, if $\zK=\zN$, then $\zx=\zK\equiv\zN$ as well. This is
because the $\zN$ distinguishable states can be chosen to be extreme,
and then they are the vertices of a $(\zN-1)$-dimensional simplex;
but this must be all of $\zSS$, since $\zSS$ is $(\zN-1)$-dimensional
(for $\zK=\zN$) and convex. (In this case, $\zRR$ is a
$\zK$-dimensional hypercube.)

\section{Maps between sets of states}
\label{sec:maps}
There is, of course, more than just one system; the transformations
and relations among systems (like the relation system-subsystem) are
extremely important.

A transformation or relation between a system with table $\zP'$ and
another with table $\zP''$ can be in some cases expressed by means of
a map $\zSS' \to \zSS''$ between the set of states $\zSS'\in
\reals^{\zK'}$ of $\zP'$, and the set of states $\zSS''\in
\reals^{\zK''}$ of $\zP''$. The map should preserve convex
combinations of state vectors\footnote{This map, however, could in
principle be \emph{partial}, \ie, not defined for all $\zs' \in
\zSS'$; this is because there may be, \eg, a transformation device
that cannot take just any state of $\zSS'$ as input.}; this implies
that its most general form is that of an \emph{affine map}, which can
be written as
\begin{equation}\label{eq:affmap}
\zs'\mapsto \zs''=\zF \zs' + \zg,
\end{equation}
where $\zF$ is a $\zK''\times \zK'$ matrix and $\zg \in
\reals^{\zK''}$ a column vector. 

It turns out, however, that such a map is always expressible as a
\emph{linear} transformation between the state vectors in $\zSS'$ and
$\zSS''$. This is due to the fact that $\zSS'$ lies in an affine
hyperplane of $\reals^{\zK'}$, and is simply proven by considering
Eq.~\eqref{eq:planeeq} for the set $\zSS'$:
\begin{equation}\label{eq:normformap}
\zn'^{\tran}\, \zs'=1, \qquad\zs'\in\zSS',
\end{equation}
where $\zn'$ is the trivial-measurement vector of $\zRR'$, and then
defining
\begin{equation}\label{eq:afftolinear}
\zc \defin \zF + \zg\zn'^{\tran},
\end{equation}
which is a $\zK''\times \zK'$ matrix representing a linear
transformation. By Eq.~\eqref{eq:normformap} one finds
\begin{equation}\label{eq:proofafftolin}
\zc \zs' = 
\zF \zs' + \zg\zn'^{\tran} \zs' =  \zF \zs' + \zg,
\end{equation}
as asserted.

Every such a map $\zc$ between states induces a dual map from the
set, $\zRR'' \in \reals^{\zK''} $, of outcomes of $\zP''$ to that,
$\zRR' \in \reals^{\zK'}$, of $\zP'$:
\begin{equation}
\zr''^{\tran} \mapsto \zr'^{\tran}=
\zr''^{\tran}\,\zc,\quad\text{or}\quad
\zr'' \mapsto \zr' = \zc^{\tran}\,\zr''.
\label{eq:dualmap}
\end{equation}
The dual map must, for obvious reasons, map the trivial-measurement
vector $\zn''$ of $\zRR''$ to the trivial-measurement vector $\zn'$
of $\zRR'$:
\begin{equation}\label{eq:constraintdualmap}
\zc^{\tran}\,\zn'' = \zn',
\end{equation}
Which represents a constraint on the possible form of $\zc$.

An analysis of the characteristics of the set of possible maps
between system states, analogous to that conducted on the sets $\zSS$
and $\zRR$, would be very interesting, but it will not be pursued
here. Such analysis could offer new perspectives (complementary to
the `tensor-product based' one) to study the relation
system-subsystem\footnote{The fact that $\zp''$ is a subsystem of
$\zp''$ can be expressed by saying that there exists a surjective,
non-injective map from $\zSS'$ to $\zSS''$. Intuitively, this is
because every preparation of a system is also a preparation for one
of its subsystems, and a preparation of a subsystem may correspond to
different preparations of the system it is part of.} and the question
of the complete positivity of
superoperators~\cite{pechukas1994,stelmachovicetal2001,peresetal2002a,peresetal2002b}.

\section{Hardy's axiomatics and quantum mechanics}
\label{sec:hardyandquantum}
The formalism just presented is quite general, containing the quantum
mechanical one as a particular case (the relation with the trace rule
is quickly shown in Appendix~\ref{sec:tracerule}), and is essentially
the same as Hardy's, prior to the introduction of his axioms. The
effect of the latter~\cite{hardy2001} is to characterise classical
and quantum systems by means of constraints on the set of states and
that of outcomes.

A classical system is characterised, in Hardy's axiomatics, by the
fact that the number of one-shot distinguishable states is maximal,
\ie, $\zN=\zK$; as a consequence, the number of extreme states is
$\zx=\zK=\zN$ as well, as shown in Sec.~\ref{sec:disting}.

A quantum mechanical system, instead, is characterised by the
supposed existence of \emph{continuous} reversible transformations
between extreme states\footnote{Hardy adopts the old saying
``\emph{natura in operationibus suis non facit saltum}'' as expressed
by
Tissot~\protect\cite{tissot1613} %\protect\nocite{fournier1857,king1904}
(see also von Linn\'e~\protect\cite{linne1751}).}. This implies that
the extreme vectors of a quantum mechanical set of states $\zSQ$ form
a continuum ($\zx=\card{\reals}$); they satisfy also a quadratic
equation which determines the ``shape'' (modulo isomorphisms, in the
convex sense) of $\zSQ$. Moreover, the maximal number of one-shot
distinguishable states is characterised by $\zK=\zN^2$. Some remarks
can be made on these features.

A continuum of extreme states can never be observed in
practice~\cite[Chap.~9]{jeffreys1931}, but is a useful approximation
or inductive generalisation which makes the powerful tools of
analysis available for doing physics. The same approximative or
inductive step is indeed taken also in classical physics (think of
classical phase space): in this case the distinguishable states, and
hence the extreme states as well, are supposed to form a continuum
($\zN=\zx=\card\reals$). From this point of view the continuum
assumption of quantum mechanics is more economical than the one of
classical physics.

On the other hand, it is to be noted that most (if not all) typical
features of quantum mechanics arise not from the continuity of the
extreme states, but from the fact that there are more extreme states
than distinguishable ones, \ie, from $\zx>\zN$ (which implies
$\zK>\zN$). This can be seen from the examples given by
Kirkpatrick~\cite{kirkpatrick2001,kirkpatrick2002,kirkpatrick2003}
(see also Ref.~\onlinecite{mana2003a}). Note, however, that in those
examples one finds that $\zK\ne\zN^2$, and $\zx$, $\zK$, $\zN$ are
finite.\footnote{It would be interesting to develop, and study the
properties of, a general mathematical formalism having no constraints
on the sets $\zSS$ and $\zRR$ or on the numbers $\zx$, $\zK$, $\zN$, and
thus containing the classical and the quantum-mechanical as special
cases.}

It is reasonable to ask, from this point of view, to what extent the
above-mentioned quantum mechanical constraints on the sets of states
and outcomes are actually observed in practice. There are cases in
which they are not; for example, in the presence of superselection
rules, where, roughly speaking, some ``portions'' of the quantum
mechanical set of states $\zSQ$ are actually missing, \ie, are not
observed (the same happens for the quantum mechanical set of outcomes
$\zRQ$).

A related question is whether, given a generic probability data table
$\zP$, the set of states $\zSS$ and the set of outcomes $\zRR$
derived from it can be ``embedded'' in some larger sets $\zSQ$ and
$\zRQ$ satisfying the quantum mechanical constraints. If this were
always possible, then quantum mechanics would be ``always right'',
just because every every experimental data table could then always be
described by quantum mechanical means.

\section{Discussion}
\label{sec:disc}
Using the conceptual tool of an imaginary `probability data table'
associated to a system, it has been shown that the states and
measurement outcomes of the system can be represented as vectors in a
real vector space. This representation simply follows from the
decomposition of the table, and in this context the rank of the
latter, $\zK$, has a peculiar r\^ole. This framework is very similar
in spirit to Hardy's framework before the introduction of his axioms,
and may thus elucidate some features of the latter.

Some properties of the sets of states and outcomes, for a generic
classical or quantum mechanical system, have then been analysed in
simple geometrical terms.

Finally, some points have been discussed concerning the
characteristics of the sets of states of a quantum mechanical system
as formalised by Hardy. In particular, it has been argued that the
origin of many typical quantum mechanical features lies not
specifically in the continuity of the extreme states, or in the
relation $\zK=\zN^2$ between $\zK$ and the number of `one-shot
distinguishable' states $\zN$, but simply in the fact that the number
of extreme states is greater than the number of `one-shot
distinguishable' states.

\begin{acknowledgments}
The author would like to thank Professor Gunnar~Bj\"ork for advice
and useful discussions, and \AA{}sa Erics\-son for a useful discussion.
\end{acknowledgments}

\appendix

\section{}
\label{sec:proofboundary}
In Sec.~\ref{sec:disting} it was stated that, if $\set{\zs_{j_1},
\dotsc, \zs_{j_{\zd}}}$ are one-shot distinguishable extreme states,
then the convex combination of any $(\zd-1)$ of them must lie on the
boundary of $\zSS$. A simple proof for three one-shot distinguishable
extreme states $\set{\zs_{j_1}, \zs_{j_2}, \zs_{j_3}}$ is the
following.

First, the condition for a point $\zs \in \zSS$ to belong to the
boundary is that there exists at least one point $\zs_{*} \in \zSS$,
$\zs_{*}\ne\zs$, such that, writing $\zs$ as a convex combination
$\zs=\zmu\zs_{*}+(1-\zmu)\zs_{**}$ ($0\le\zmu\le1$) of $\zs_{*}$ and some
other point $\zs_{**}$, implies that $\zmu=0$ (\ie, the combination
cannot be \emph{proper}). Intuitively, this means that straight lines
cannot be drawn from the point $\zs$ in just any direction, if they
are to remain inside $\zSS$.

Now, consider a point $\zs$ given by a convex combination of
$\zs_{j_1}$ and $\zs_{j_2}$:
\begin{equation}
\label{eq:convcomb1}
\zs\equiv \zla \zs_{j_1}+ (1-\zla) \zs_{j_2},\qquad 0\le\zla\le1.
\end{equation}
since $\zs_{j_1}$, $\zs_{j_2}$, and $\zs_{j_3}$ are one-shot
distinguishable, there must exist a measurement outcome $\zr$ such
that
\begin{align}
\zr \inn \zs_{j_1}&=\zr \inn \zs_{j_2}=0,\label{eq:dist0}\\
\zr \inn \zs_{j_3}&=1,\label{eq:dist1}
\end{align}
and Eqs.~\eqref{eq:convcomb1},~\eqref{eq:dist0} yield
\begin{equation}
\label{eq:combiszero}
\zr\inn \zs=\zla \zr \inn \zs_{j_1}+ (1-\zla) \zr \inn \zs_{j_2} =0.
\end{equation}
Now write $\zs$ as a convex combination of the vector $\zs_{j_3}$
with some other vector $\zs_{**}$:
\begin{equation}
\label{eq:hypotcomb}
\zs=\zmu \zs_{j_3}+ (1-\zmu) \zs_{**},\qquad 0\le\zmu\le1.
\end{equation}
From Eqs.~\eqref{eq:dist0}, \eqref{eq:combiszero},
and~\eqref{eq:dist1}, one obtains
\begin{equation}
\label{eq:nonposscomb}
0=\zr \inn \zs= \zmu \zr \inn \zs_{j_3}+ (1-\zmu) \zr \inn\zs_{**}
=\zmu + (1-\zmu) \zr \inn\zs_{**} 
\end{equation}
which can be satisfied only if $\zmu=0$ (which implies
$\zs_{**}=\zs$). Thus, the vector $\zs_{j_3}$ plays the role of
$\zs_{*}$ in the condition given above, and so any $\zs =\zla
\zs_{j_1}+ (1-\zla) \zs_{j_2}$ lies in the boundary of $\zSS$. An
analogous proof holds for any convex combination of any two of the
vectors $\set{\zs_{j_1}, \zs_{j_2}, \zs_{j_3}}$, and this implies
that the latter are vertices of a triangle which is part of the
boundary of $\zSS$.

The generalisation to more than three vectors is straightforward.

\section{The trace rule}
\label{sec:tracerule}
It is shown that the `scalar product formula',
Eq.~\eqref{eq:baserule}, includes also the `trace rule' of quantum
mechanics (\cf\ Hardy~\cite[Sec.~5]{hardy2001}; also
Weigert~\cite{weigert2000}).

A state is usually represented in quantum mechanics by a
\emph{density matrix} $\zrho_j$, and a measurement outcome by a
\emph{positive-operator-valued measure element} $\zpov_i$; both are Hermitian
operators in a Hilbert space of dimension $\zN$. The probability of
obtaining the outcome $\zpov_i$ for a given measurement on state
$\zrho_j$ is given by the trace formula
\begin{equation}
\zp_{ij}=\tr\zpov_i \zrho_j.\label{eq:traceform}
\end{equation}
The Hermitian operators form a linear space of \emph{real} dimension
$\zK=\zN^2$; one can choose $\zN^2$ linearly independent Hermitian
operators $\set{\zbas_k}$ as a basis for this linear space. These can
also be chosen (basically by Gram-Schmidt orthonormalisation) to
satisfy
\begin{equation}
\label{eq:basiskron}
\tr{\zbas_k \zbas_l}=\kron_{kl}.
\end{equation}
Both $\zrho_j$ and $\zpov_i$ can be written as a linear combination of
the basis states
\begin{equation}
\label{eq:writewithbasis}
\zrho_j=\sum_{l=1}^{\zK} {\zS_j}^l \zbas_l,\qquad
\zpov_i=\sum_{k=1}^{\zK} {\zR_i}^k \zbas_k,
\end{equation}
where the coefficients ${\zS_j}^l$ and ${\zR_i}^k$ are real. Using
Eqs.~\eqref{eq:writewithbasis} and~\eqref{eq:basiskron} the trace
formula becomes
\begin{equation}
\zp_{ij}=\tr\zpov_i \zrho_j=\sum_{k,l=1}^{\zK} {\zR_i}^k {\zS_j}^l
\tr\zbas_k \zbas_l=
\sum_{l=1}^{\zK} {\zR_i}^l {\zS_j}^l = 
\zr_i \inn\zs_j,\label{eq:scalarprform}
\end{equation}
where $\zr_i\defin \bigl({\zR_i}^1\dotso{\zR_i}^{\zK}\bigr)$ and
$\zs_j\defin \bigl({\zS_j}^1\dotso{\zS_j}^{\zK}\bigr)$ are vectors in
$\reals^{\zK}$.

%\bibliography{bib}

\end{document}